\documentstyle[twoside,fleqn,espcrc2,amssymb,epsfig]{article}
\title{Probing the parton densities of polarized photons 
at a linear $e^+e^-$ collider}

\author{M.\ Stratmann\address{Department of Physics, University of Durham,
                              Durham DH1 3LE, England}}

\begin{document}

\begin{abstract}
The present theoretical status of spin-dependent parton densities 
$\Delta f^{\gamma}(x,Q^2)$ of circularly polarized photons is
briefly reviewed.
It is then demonstrated that measurements of the deep-inelastic spin asymmetry 
$A_1^{\gamma}\simeq g_1^{\gamma}/F_1^{\gamma}$ and of di-jet 
rapidity distributions at a future linear $e^+e^-$ collider appear 
to be particularly suited for a determination 
the spin-dependent photonic quark and gluon densities, respectively.
Special emphasis is devoted to a comparison of the different sources 
of polarized  photons at a linear collider: the equivalent photon
approximation and backscattered laser (Compton) photons.
It is shown that backscattered laser photons are highly favorable,
even indispensable, for decent measurements of the $\Delta f^{\gamma}(x,Q^2)$.
\end{abstract}

\maketitle

\section{$\boldmath{\Delta f^{\gamma}}$: FORMALISM AND PRESENT STATUS}

Recent months have seen a substantial amount of new 
experimental results on {\em unpolarized} 
deep-inelastic electron-photon scattering from LEP \cite{ref:lepdata} which
provide a considerably extended kinematical coverage in
$x$ and $Q^2$ as compared to all previous results since PEP and PETRA.
Complementary information on the partonic structure
of photons is provided by photoproduction measurements at HERA, 
in particular from (di-)jet production data \cite{ref:heradata}, 
and combining these
results should not only vastly improve our knowledge 
of the photon structure but also seriously challenges the 
presently available theoretical models.

A similar analysis in {\em longitudinally polarized} $e^+e^-$ and $ep$ 
collisions would be desirable.
Measuring the difference between the independent helicity
combinations of the incoming particles, 
\begin{equation}
\label{eq:xsecdef}
\Delta\sigma = \frac{1}{2}\left[ \sigma(++) - \sigma(+-) \right]\;\;,
\end{equation}
instead of the sum, as in unpolarized (helicity-averaged) experiments,
would give access to the parton structure of circularly 
polarized photons, which is completely unknown experimentally so far. 
These distributions are defined by
\begin{equation}
\label{eq:pdfdef}
\Delta f^{\gamma}(x,Q^2) \equiv f_+^{\gamma_{+}}(x,Q^2) -  
f_-^{\gamma_{+}}(x,Q^2)\;\;,
\end{equation}
where $f_+^{\gamma_{+}}$ $(f_-^{\gamma_{+}})$ 
denotes the density of a parton $f$ with helicity `+' (`$-$') 
in a photon with helicity `+'. 
The densities $\Delta f^{\gamma}$ contain information different
from that contained in the unpolarized ones [defined
by taking the sum in (\ref{eq:pdfdef})], and their measurement is
necessary for a thorough understanding of the partonic structure of photons.

The complete next-to-leading order (NLO) QCD framework 
for the $Q^2$-evolution of $\Delta f^{\gamma}$ 
and the calculation of the polarized photon
structure function $g_1^{\gamma}$ was recently provided in 
\cite{ref:nloletter} and will be briefly reviewed here 
(see also \cite{ref:lund}).
The $\Delta f^{\gamma}$ obey the well-known {\em inhomogeneous} evolution 
equations schematically given by 
\begin{equation}
\label{eq:gl1}
\frac{d \Delta q_i^{\gamma}}{d \ln Q^2} = \Delta k_i +
\left( \Delta P_i \otimes \Delta q_i^{\gamma} \right) 
\end{equation}
(any obvious $x$ and $Q^2$ dependence is suppressed here and 
in what follows), where $q_i^{\gamma}$ stands for the flavor non-singlet 
(NS) quark combinations or the singlet (S) vector 
$\Delta \vec{q}^{\,\gamma}_{S} \equiv {\Delta \Sigma^{\gamma}
\choose \Delta g^{\gamma}}$. The symbol $\otimes$ denotes the usual 
convolution in Bjorken-$x$ space. 
The evolution equations (\ref{eq:gl1}) are most conveniently solved in
Mellin-$n$ moment space, where the solutions can be given analytically.
The 1-loop (LO) and 2-loop (NLO) contributions to
polarized photon-to-parton and parton-to-parton splitting functions,
$\Delta k_i$ and $\Delta P_i$, in (\ref{eq:gl1}) can be 
found in \cite{ref:nloletter} and \cite{ref:splfct}, respectively.
The solution of (\ref{eq:gl1}) can be decomposed into a `pointlike' 
(inhomogeneous) and a `hadronic' (homogeneous) part, 
\begin{equation}
\label{eq:solution}
\Delta q_i^{\gamma} = \Delta q^{\gamma}_{i,PL} +
\Delta q^{\gamma}_{i,had}
\end{equation}
($i=$ NS, S) and can be found in \cite{ref:disgamma} 
(with obvious replacements of all unpolarized quantities by their 
polarized counterparts).

Spin-dependent deep-inelastic electron-photon scattering (DIS) can be 
parametrized in terms of a {\em new}, polarized structure function 
$g_1^{\gamma}(x,Q^2)$, in analogy to $F_2^{\gamma}$ and $F_L^{\gamma}$ 
in the helicity-averaged case, and the NLO expression for 
$g_1^{\gamma}$ reads \cite{ref:nloletter} 
\begin{eqnarray}
\label{eq:g1}
\nonumber
g_1^{\gamma} \!\!\!&=&\!\!\!  \frac{1}{2} \sum_{f=u,d,s} \!e_f^2\;
\Bigg\{ 2 \Delta f^{\gamma} 
+ \frac{\alpha_s}{2\pi} \bigg[ 2 \Delta C_q \otimes 
\Delta f^{\gamma} \\
\!\!\!&+&\!\!\!
\frac{1}{N_f} \Delta C_g \otimes \Delta g^{\gamma}\bigg] \Bigg\}  
+ \frac{3\alpha}{4\pi} \sum_{f=u,d,s} \!e_f^4 \Delta C_{\gamma}\,.
\end{eqnarray}
The coefficient functions $\Delta C_q$ and $\Delta C_g$
can be found in the $\overline{\rm{MS}}$ scheme, e.g., 
in \cite{ref:splfct}, and $N_f$ being the number of active flavors.
The $\ln (1-x)$ dependence of the photonic coefficient $\Delta C_{\gamma}$ 
\cite{ref:nloletter} causes perturbative instability problems
for $x\rightarrow 1$ in the  $\overline{\rm{MS}}$ scheme.
But, as in the unpolarized case \cite{ref:disgamma}, one can overcome 
this `problem' by absorbing $\Delta C_{\gamma}$ into the definition 
of the quark densities \cite{ref:nloletter} ($\rm{DIS}_{\gamma}$ scheme).
Only the contribution of the light flavors has been written out in  
(\ref{eq:g1}). Heavy quark contributions 
to $g_1^{\gamma}$ should be more appropriately included via the 
relevant fully massive polarized boson fusion subprocesses 
(see, e.g., \cite{ref:gsv}).

At present one has to fully rely on theoretical models for the 
$\Delta f^{\gamma}$.
The only guidance is provided by the positivity of the helicity dependent
cross sections on the r.h.s.\ of (\ref{eq:xsecdef}) demanding that
$|\Delta \sigma| \le \sigma$. This can be directly translated into
a useful constraint on the densities\footnote{Strictly speaking 
positivity applies only to physical quantities like cross sections and not to
parton densities beyond the LO where they become scheme-dependent 
(`unphysical') objects. Of course, (\ref{eq:pos}) still serves as a
reasonable `starting point' for the NLO densities.} in terms of the
already known unpolarized distributions $f^{\gamma}$:
\begin{equation}
\label{eq:pos}
|\Delta f^{\gamma}(x,Q^2)| \le f^{\gamma}(x,Q^2)\;\;.
\end{equation}

In addition, it was shown \cite{ref:currentcons} that the first moment 
$(n=1)$ of $g_1^{\gamma}$ vanishes irrespective of 
$Q^2$ [`current conservation' (CC)].
This result holds to all orders in perturbation theory and at every twist 
provided that the fermions in the theory have non-vanishing mass 
\cite{ref:currentcons}.
Due to the properties of the splitting and coefficient functions for $n=1$,
CC is automatically fulfilled for the pointlike part
of $g_1^{\gamma}$ and hence can be enforced for the full $g_1^{\gamma}$ by
demanding a vanishing hadronic quark input for $n=1$ 
(the gluon is not constrained in the $\overline{\mathrm{MS}}$ or
$\mathrm{DIS}_\gamma$ scheme since $\Delta C_g=0$ for $n=1$).

To obtain a realistic estimate for the theoretical uncertainties in 
$\Delta f^{\gamma}$ coming from the unknown hadronic input, 
we consider two very different scenarios in LO\footnote{In what follows we
limit ourselves to LO which is sufficient to estimate the
sensitivity of future experiments to the unknown $\Delta f^{\gamma}$.
Both scenarios can be straightforwardly extended to NLO, 
see \cite{ref:nloletter} for details.} \cite{ref:gv,ref:gsv} based 
on the positivity bound (\ref{eq:pos}):
for the first (`{\em maximal scenario}') we saturate (\ref{eq:pos})
using the phenomenologically successful unpolarized GRV photon densities
\cite{ref:grvphot} 
\begin{equation}
\label{eq:maxsat} 
\Delta f^{\gamma}_{had}(x,\mu^2) = f_{had}^{\gamma}(x,\mu^2)\;\;,
\end{equation}
whereas the other extreme input (`{\em minimal scenario}') is defined by
\begin{equation}
\label{eq:minsat}
\Delta f^{\gamma}_{had}(x,\mu^2) = 0
\end{equation}
with $\mu\simeq 0.6\,{\rm{GeV}}$ \cite{ref:grvphot}.
Of the two extreme hadronic inputs only (\ref{eq:minsat}) satisfies CC.
However, if one is interested only in the region of, say,
$x\gtrsim 0.005$, CC could well be implemented by contributions 
from smaller $x$, which do not affect the evolutions at larger $x$. 
Rather than artificially enforcing the vanishing of the first moment 
of the $\Delta q_{had}^{\gamma} (x,\mu^2)$ in (\ref{eq:maxsat}),
we therefore stick to the two extreme scenarios as introduced above.

\begin{figure*}[tbh]
\begin{center}
\vspace*{-0.6cm}
\epsfig{file=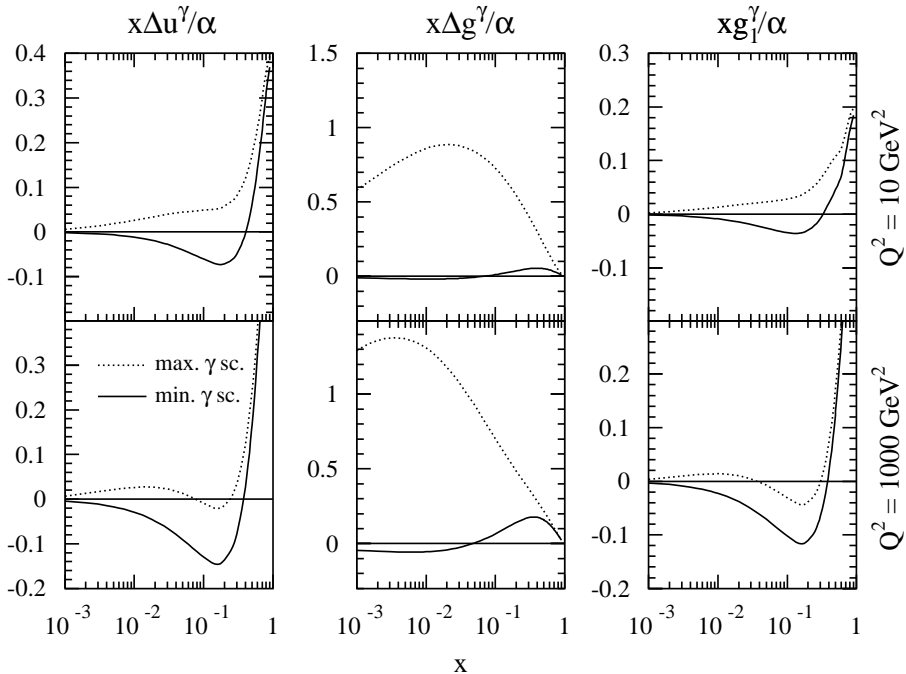,width=13cm}
\end{center}
\vspace*{-1.4cm}
\caption{$xu^{\gamma}/\alpha$ and $xg^{\gamma}/\alpha$  
evolved to $Q^2=10$ and $1000\,\mathrm{GeV}^2$ in LO using the two 
extreme inputs (\ref{eq:maxsat}) and (\ref{eq:minsat}).
Also shown is the structure function $g_1^{\gamma}$ in LO.}
\vspace*{-0.3cm}
\end{figure*}
Figure 1 compares our LO $u$-quark and gluon distributions and the structure
function $g_1^{\gamma}$ for the two extreme scenarios based on
(\ref{eq:maxsat}) and (\ref{eq:minsat}) 
at $Q^2=10$ and $1000\,\mathrm{GeV}^2$.
It is interesting to note that the pointlike evolution leads 
to a sizeable negative $u$-quark distribution for $x$ around $0.1$ 
in the `minimal' scenario despite of the vanishing input (\ref{eq:minsat}). 
In the shown, experimentally relevant $Q^2$ region the quark densities 
in the `maximal' scenario mainly have a different sign 
and are smaller in their absolute size. Hence
one should expect spin asymmetries with opposite signs for quark 
dominated processes like DIS (cf.\ the results for $g_1^{\gamma}$ in Fig.~1 
and Fig.~3).
On the contrary the gluon distribution remains small in the
`minimal' scenario for all values of $Q^2$, and one thus expects almost
vanishing polarized cross sections and spin asymmetries for gluon dominated
processes such as jet production (cf.\ Fig.~4 below).  

To finish this section let us briefly turn to the two conceivable future
experiments: running HERA in a polarized collider mode at some stage 
in the future \cite{ref:polhera} or using a linear $e^+e^-$ collider 
with polarized beams.
Measurements of spin asymmetries in, e.g., the photoproduction of 
large-$p_T$ (di-)jets can then in principle reveal information on the 
$\Delta f^{\gamma}$ through the presence of `resolved' photon 
processes in the first case. This has been extensively studied in recent years 
and was shown to be feasible \cite{ref:heraresults1,ref:heraresults2}. 
Here we focus exclusively 
on the latter option which has not attracted much interest so far. 
A future polarized linear $e^+e^-$ collider can provide complementary 
information on the $\Delta f^{\gamma}$ by measuring the
structure function $g_1^{\gamma}(x,Q^2)$ or
spin asymmetries in resolved two-photon reactions \cite{ref:svlinear}. 

\section{POLARIZED PHOTON SPECTRA}
%
Before turning to a study of possible tests and signatures for
$\Delta f^{\gamma}$ it is worthwhile and important to explore and compare 
the different sources for polarized photons at a linear collider first.
This is because all experimentally accessible spin asymmetries\footnote{All
spin experiments measure spin asymmetries rather than $\Delta \sigma$
directly because normalization uncertainties, etc., conveniently drop out in 
the cross section ratio.} can be {\em roughly} decomposed as
\begin{equation}
\label{eq:decomp}
A\equiv \frac{\Delta \sigma}{\sigma} 
\approx A_{\mathrm{flux}} \times A_{f^{\gamma}} \times A_{\hat{\sigma}}\;\;,
\end{equation}
where $(\Delta) \sigma$ denotes the (polarized) unpolarized cross section
under consideration, $A_{\mathrm{flux}}$ is the yield of polarized photons, 
$A_{f^{\gamma}}$ is related to certain combinations of 
$\Delta f^{\gamma}/f^{\gamma}$,
and $A_{\hat{\sigma}}\equiv \Delta \hat{\sigma}/ \hat{\sigma}$ is the partonic
subprocess cross section asymmetry.
The latter quantity is calculable in perturbative QCD, while $A_{f^{\gamma}}$
is sensitive to the unknown distributions $\Delta f^{\gamma}$ 
we are looking for.
Obviously, the photon flux spin asymmetry in (\ref{eq:decomp}) should be as
large as possible to make an extraction of $\Delta f^{\gamma}$ feasible, 
ideally $A_{\mathrm{flux}}\simeq 1$ largely independent of 
different kinematical configurations.

The `standard' description for polarized photons is
the equivalent photon approximation (EPA)
where a circularly polarized photon is 
collinearly radiated off a longitudinally polarized electron.
The EPA spectrum is shown in Fig.~2(a) for a typical set of parameters 
at a future linear collider, and the flux spin asymmetry is approximately 
given by the electron-to-photon splitting function ratio 
\begin{equation}
\label{eq:ww}
A_{\mathrm{flux}} \simeq \frac{1-(1-y)^2}{1+(1-y)^2}\;\;,
\end{equation}
where $y$ is the energy fraction of the photon taken away from the
electron. As can be seen from (\ref{eq:ww}) and Fig.~2(a), the
flux asymmetry becomes rapidly less and less favorable as $y\rightarrow 0$.
In addition the spectrum is broad, i.e., not mono-energetic, which causes
experimental problems in DIS since the $e\gamma$ c.m.s.~energy 
is not known and the momentum fraction $x$ cannot be reconstructed 
as in $ep$ scattering.

\begin{figure}[tbh]
\vspace*{-0.7cm}
\epsfig{file=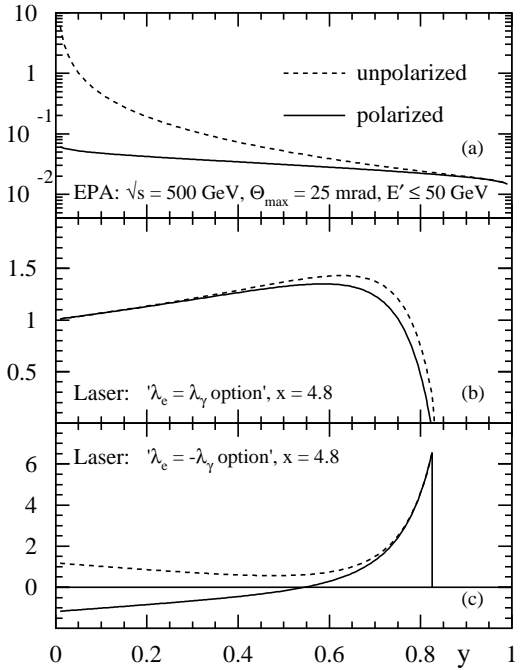,width=8.5cm}
\vspace*{-1.9cm}
\caption{Comparison of the unpolarized and polarized photon spectra as a 
function of the photon's energy fraction $y$ for (a): EPA, 
(b): Compton backscattering for like-handed
electrons and laser photons, and (c): as (b) but for opposite helicities.}
\vspace*{-0.5cm}
\end{figure}
Photon beams can be obtained also via Compton backscattering of laser
photons off the linac electrons \cite{ref:laser1}. These $\gamma\gamma$
colliders, which are experimentally challenging and have not been realized
yet, were so far mainly regarded as a novel tool to study, e.g.,  properties 
of intermediate mass Higgs bosons, but they are also particularly 
suited for QCD studies of the (un)polarized photon structure.
Polarizing the electron beams and/or laser photons not only provides 
circularly polarized backscattered photons but also allows one to tailor the
energy spectrum of the photon to one's needs. Colliding like-handed electrons 
and laser photons results in a broad spectrum and colliding oppositely
handed electrons and photons results in a peaked distribution for the
backscattered photon as can be seen in Figs.~2(b) and (c), respectively.
It is important to notice that in both cases the resulting photons
are {\em highly polarized}, i.e., $\left|A_{\mathrm{flux}}\right|\simeq 1$
in (\ref{eq:decomp}). The helicity dependent cross sections for 
such a $\gamma\gamma$ collider have to be labeled
by four helicities, the helicities of both the two electron beams and the
two laser photons, instead of two as on the r.h.s.\ of (\ref{eq:xsecdef}).
This allows, of course, more ways to define $\Delta \sigma$, 
however, in practice the use of either of the two spectra shown in Fig.~2 
seems to be the most promising option. 
Apart from the helicities the Compton kinematics is controlled by
a dimensionless variable $x$ which is correlated with the maximum energy
of the backscattered photons. 
In Fig.~2 we have chosen the maximum value $x=4.8$
which just prevents reconversion of the high-energetic photon into an
$e^+e^-$ pair \cite{ref:laser1}.

\section{$\boldmath{\Delta f^{\gamma}}$:  TESTS AND SIGNATURES}
%
Equipped with the technical framework let us now study possible tests
and signatures for the $\Delta f^{\gamma}$ at a future linear
$e^+e^-$ collider with a c.m.s.~energy of $\sqrt{s}=500\;\mathrm{GeV}$
and choice of either having the EPA or backscattered laser photons 
as the source for polarized photons.

\begin{figure*}[tbh]
\begin{center}
\vspace*{-1.1cm}
\epsfig{file=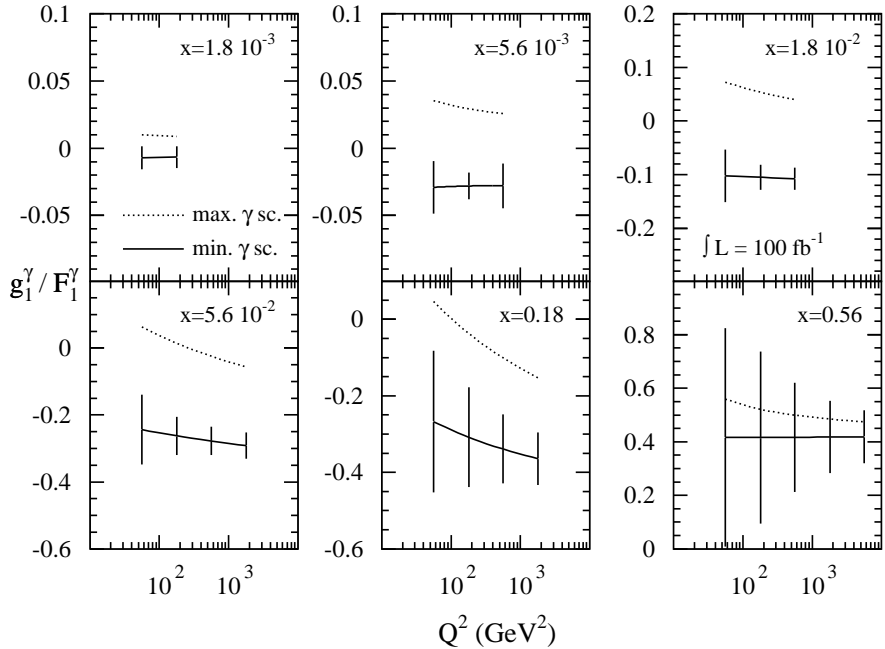,width=13cm}
\end{center}
\vspace*{-1.35cm}
\caption{Predictions for the DIS spin asymmetry $g_1^{\gamma}/F_1^{\gamma}$
in LO using the backscattered laser photon spectrum according to 
Fig.~2(b) and the two scenarios for $\Delta f^{\gamma}$ as
outlined above. The unpolarized structure function $F_1^{\gamma}$ was
calculated using the LO GRV densities \cite{ref:grvphot}.
The error bars denote the expected statistical uncertainty for such a
measurement assuming an integrated luminosity of $100\,\mathrm{fb^{-1}}$,
$70\%$ beam polarization, and two bins in $x$ and $Q^2$ per decade.}
\vspace*{-0.4cm}
\end{figure*}
Figure 3 shows predictions for the DIS spin asymmetry 
$A_1^{\gamma}\simeq g_1^{\gamma}/F_1^{\gamma}$ in LO using the 
two extreme models for the $\Delta f^{\gamma}$ based on the inputs 
(\ref{eq:maxsat}) and (\ref{eq:minsat}).
For the target photon, the photon flux of Fig.~2(b) was 
used\footnote{For simplicity we ignore here experimental problems
due to the unknown precise value of the $e\gamma$ c.m.s.~energy
caused by the tail to low $y$ in the photon flux.
The peaked spectrum in Fig.~2(c) also has a tail which has
to be cut-off somehow. Specially tailored spectra should lead, however, 
to qualitatively very similar results as in Fig.~3.}.
Also shown is the expected statistical uncertainty for such a
measurement assuming an integrated luminosity of 
${\cal{L}}=100\,\mathrm{fb^{-1}}$ and two bins in $x$ and $Q^2$ per decade.
Given these error bars it should be feasible to distinguish between the
two extreme scenarios at not too large values of $x$. For $x\rightarrow 1$
the two models become very similar anyway due to the dominance of the
pointlike part.
It is important to stress that a study of $g_1^{\gamma}/F_1^{\gamma}$
appears to be {\em not} feasible without having backscattered laser
photons as target because the statistical errors would become 
too large.

Since (di)-jet production was shown to be a very promising tool
to decipher the $\Delta f^{\gamma}$ at a future polarized HERA $ep$
collider \cite{ref:heraresults1,ref:heraresults2}, we extend 
these studies here to linear $e^+e^-$ colliders. 
Di-jet production is particularly suited as it allows to define 
single (double) resolved photon samples on an experimental basis by 
demanding that the momentum fractions for one (both) photon(s) are less than
some cut, say, $x_{\gamma}^{\pm}\leq 0.8$. 
The direct photon `background', 
$\gamma\gamma\rightarrow q\bar{q}$, which is concentrated 
at $x_{\gamma}^{\pm}\simeq 1$ is discarded in that
way. In the unpolarized case such measurements were performed
at LEP2 by OPAL \cite{ref:opal} recently.

\begin{figure}[tbh]
\vspace*{-0.2cm}
\epsfig{file=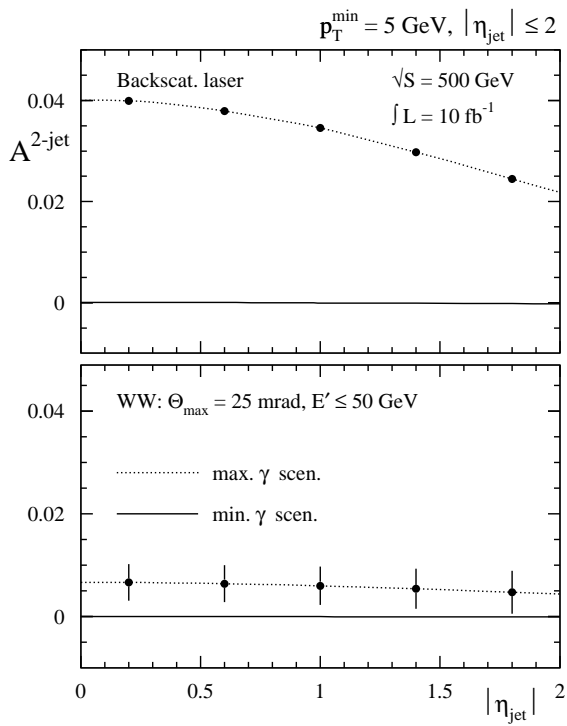,width=8.5cm}
\vspace*{-1.8cm}
\caption{Double resolved $(x_{\gamma}^{\pm}\leq 0.8)$
di-jet spin asymmetry $A^{\mathrm{2-jet}}$ in LO.
The unpolarized cross sections were calculated using the GRV
densities \cite{ref:grvphot}.
For backscattered laser photons the spectrum in Fig.~2(b) was used, 
while in the lower part the EPA was used as in Fig.~2(a). 
The statistical error bars assume a luminosity of $10\,\mathrm{fb}^{-1}$
and $70\%$ beam polarization.}
\vspace*{-0.5cm}
\end{figure}
Fig.~4 shows our expectations for the double resolved
di-jet spin asymmetry $A^{\mathrm{2-jet}}$ in LO using similar cuts in
transverse momentum $p_T$ and rapidity $\eta_{\mathrm{jet}}$ 
of the jets as in \cite{ref:opal}.
$A^{\mathrm{2-jet}}$ is dominated by gluon-gluon initiated subprocesses,
and hence an almost vanishing asymmetry is obtained
for the minimal scenario as expected from Fig.~1.
Again, backscattered Compton photons are strongly favored
compared to the equivalent photon spectrum, but a measurement 
might be possible also in the latter case here. 

It turns out that uncertainties due to the choice of the factorization
scale, which are expected to be particularly pronounced in LO, 
cancel to a large extent in the ratio $A^{\mathrm{2-jet}}$.
For jet production in polarized $ep$ collisions it 
was shown that LO QCD agrees reasonably well with Monte Carlo (MC)
results including parton showering and hadronization \cite{ref:heraresults2},
however, due to the lack of a MC generator for 
polarized $\gamma\gamma$ collisions such a study is not possible here
for the time being.

To actually unfold information on $\Delta f^{\gamma}$ 
it is useful to introduce the concept of `effective parton
densities' \cite{ref:effpdf}. Although  $A^{\mathrm{2-jet}}$ 
is dominated by $gg$ scattering, all
QCD subprocesses contribute, which considerably complicates any analysis. 
In the unpolarized case it was shown \cite{ref:effpdf}
that the ratios of dominant subprocesses, those with a $t$-channel gluon
exchange, are roughly constant, i.e., $qq'/qg\simeq qg/gg\simeq 4/9$. 
Hence by introducing a suitable effective parton density 
$f_{\mathrm{eff}}=\sum_q (q+\bar{q})+9g/4$ the jet cross section 
factorizes into these densities times a {\em single} subprocess 
cross section.
In the polarized case this factorization is slightly broken as
$qq'/qg \neq qg/gg$. However, the approximation still
works surprisingly well at a level of $5-10\%$ accuracy, and 
the appropriate effective densities are given by \cite{ref:svlinear}
\begin{equation}
\label{eq:effpdf}
\Delta f_{\mathrm{eff}}^{\gamma} = \sum_q (\Delta q^{\gamma}
                                   +\Delta \bar{q}^{\gamma})+
                                   \frac{11}{4} \Delta g^{\gamma}
\end{equation} 
such that the polarized double resolved jet cross section can be
expressed as
\begin{equation}
\label{eq:effxsec}
\Delta \sigma^{\mathrm{2-jet}} \simeq \Delta f_{\mathrm{eff}}^{\gamma} \otimes
                     \Delta f_{\mathrm{eff}}^{\gamma} \otimes
                     \Delta \hat{\sigma}_{qq'\rightarrow qq'}\,\,.
\end{equation}
Having extracted $\Delta f_{\mathrm{eff}}^{\gamma}$ from a
measurement of $A^{\mathrm{2-jet}}$ exploiting (\ref{eq:effxsec}) 
and $\Delta q^{\gamma}$ from
elsewhere, e.g., from DIS, one can determine $\Delta g^{\gamma}$ 
using (\ref{eq:effpdf}).

\section{CONCLUDING REMARKS}
%
It was shown that a future linear $e^+e^-$ collider could reveal {\em first}
information on the presently completely unmeasured parton densities
of circularly polarized photons. Studies of the spin asymmetry in
DIS and double resolved di-jet production appear to be particularly 
suited. It was demonstrated that it would be highly favorable, even
mandatory in case of DIS, to have backscattered laser photons available
for these measurements. Compared to a measurement of the $\Delta f^{\gamma}$
in resolved photoproduction processes at a polarized HERA, a linear
collider would provide a much `cleaner' environment since the 
spin-dependent proton densities do not enter here and thus
do not interfere with the extraction of $\Delta f^{\gamma}$.

\section*{ACKNOWLEDGMENTS}
%
It is a pleasure to thank W.\ Vogelsang for a pleasant collaboration
on all topics presented here.


%
\end{document}